\documentclass[a4paper,11pt]{article}
\usepackage{jinstpub}
\usepackage{lineno}
\usepackage[utf8]{inputenc}
\usepackage{amsmath}
\usepackage{upgreek}
\usepackage{ulem}

\title{\boldmath Analysis of the S1 triplet component in the DarkSide-50 experiment}

\author[a]{Clea Sunny\footnote{Speaker}}
\collaboration[c]{on behalf of DarkSide-50 collaboration}
\affiliation[a]{AstroCeNT, Centrum Astronomiczne im. Miko\l{}aja Kopernika Polskiej Akademii Nauk,\\
Rektorska 4, 00-614 Warszawa, Poland}

\emailAdd{sclea@camk.edu.pl}

\abstract{The DarkSide-50 (DS-50) experiment aims at the direct detection of weakly interacting massive particles.  It is a dual-phase liquid argon time projection chamber (LAr TPC) where Dark Matter (DM), which constitutes five sixths of all matter in the universe, is expected to interact with an argon nucleus resulting in a nuclear recoil. A scintillation signal (S1) is produced as a result of the ionising events from the DM-Ar interaction.  The impurities in LAr, such as $\text{O}_2, \text{N}_2, \text{H}_2\text{O}$, etc., at the ppm level, quench the scintillation photons, leading to a reduction in the observed lifetime of the triplet state. In this contribution, the effect of impurities on the triplet lifetime is analyzed for individual events using DS-50 data, with a primary focus on nitrogen, one of the candidates for the impurities in LAr hypothesized to cause a suppression of triplet lifetime. This is done by determining the lifetime of the triplet component with a known purity, which can be used as a reference for the purity level of argon used in current and future dark matter searches. }


\begin{document}
\maketitle
\flushbottom

\section{Introduction}

Weakly Interacting Massive Particles or WIMPs are considered as one of the most promising DM candidates. Numerous experiments have been carried out in order to search for WIMP signals in various mass ranges. DS-50 is one such experiment which probed for WIMPs in both high-mass (>10 GeV/\(c^2\)) and low-mass (<10 GeV/\(c^2\)) ranges \cite{a}. The DS-50 program uses a Liquid Argon Time Projection Chamber (LAr TPC), where both DM-induced Nuclear Recoil (NR) and background-induced Electron Recoil (ER) signals can be seen. 

In the DS-50 experiment, the argon gas drawn from the cryostat is constantly circulated through a loop in the gas phase containing a hot getter (SAES Monotorr PS4-MT50-R-2) for its purification \cite{b}. The impurity molecules are irreversibly bound to the materials of the getter, thereby removing impurities to below 1 ppb level. When the getter is not in use, impurities outgassing from the detector materials may accumulate in the argon gas. 

A dedicated research has been conducted recently with DS-50 data on Spurious Electrons (SEs), which are isolated ionization electrons extracted from the LAr that reach the gas layer above the LAr surface. Analysis of the low-energy backgrounds in DS-50 has encountered a surplus of SE signals, when the getter was bypassed. The cause of SEs is yet to be determined, but one possibility is that these SEs are caused by the presence of impurities in LAr during the getter-off period. From previous studies on the contamination in LAr, it has been noted that contaminants such as $\text{O}_2, \text{N}_2, \text{and } \text{H}_2\text{O},$ with concentrations at the order of parts per million (ppm) can influence the argon scintillation, due to the quenching processes that occur between the impurity molecules and the $\text{Ar}_2$ excimer states \cite{c,d}. Several mechanisms could produce SEs, with one hypothesis suggesting that these are re-emitted electrons from metastable anions formed when impurities capture drifting electrons. The highly electronegative $\text{O}_2$ is a likely candidate for the impurity, but the electron lifetime in DS-50, which is the time before a drifting ionization electron attaches to an impurity, showed no degradation during the getter-off period, indicating that $\text{O}_2$ did not accumulate at these times. Therefore, impurities causing the SE signals must capture electrons at a lesser rate. This is the reason why $\text{O}_2$ is not taken into account in this analysis.

The SE rate when the getter was bypassed is found to be $(3.5 \pm 0.3) \times 10^{-5}$ e$^-$/e$^-$, and that after the getter was re-installed is measured as $(0.5 \pm 0.1) \times 10^{-5}$ e$^-$/e$^-$ normalized to the total yield of ionization electrons \cite{e}. The observed faster reduction in the SE rate after the getter was re-installed, compared to the getter-off period, may indicate that if SEs are caused by an impurity, then the impurity is removable by the gas purification system. This may imply that the cause of SEs could be an impurity that predominantly exists in the gas phase. Given that $\text{N}_2$ is likely volatile at 87 K and captures electrons with a lower electron affinity in LAr than $\text{O}_2$, $\text{N}_2$ is considered one of the prime candidates.

This analysis aims at testing the hypothesis of $\text{N}_2$ being the cause of SEs, by examining the triplet component of argon scintillation. It is well known that $\text{N}_2$ at the level of ppm can reduce the lifetime of the triplet component and the overall scintillation light yield \cite{c}.

In this study, using the UAr data of DS-50, acquired between April 2015 and April 2018, the lifetime of the triplet component is determined for three years. A significant degradation in the triplet lifetime is expected at times when the purification system was not used, indicating the presence of impurities in the LAr at these times. When inferring impurity levels from the triplet lifetime, the assumption is made that $\text{N}_2$ is the impurity to blame.

\section{Scintillation light emission in the DS-50 experiment}

The DS-50 detector, operating in the underground Laboratory Nazionali del Gran Sasso (LNGS), has a double-phase LAr TPC filled with 50 kg of liquefied Underground Argon (UAr), which has a reduced concentration of $^{39}\text{Ar}$ in the active volume. The activity of $^{39}\text{Ar}$ in DS-50 is measured to be $0.73$ mBq/kg \cite{f}. The signals are recorded using 38 Hamamatsu $\text{R}110653$ low-background, high-quantum-eﬃciency Photo Multiplier Tubes (PMTs), 19 both on the top and the bottom of the active volume inside the detector. The DS-50 data was taken with a drift field of 200 V/cm, and an extraction field of 2.8 kV/cm \cite{g}. 

An event in DS-50 consists of two main signals: a scintillation signal, S1, produced as the result of the interaction between an Ar atom and an incoming particle in LAr, and an ionization signal, S2, resulting from the excitation of gaseous argon atoms in the argon gas layer between the LAr surface and the TPC anode. This excitation is caused by the ionization electrons, which are drifted upwards within the LAr volume, escaping recombination in the TPC, and then are extracted and accelerated by the electric field. The photoelectron (PE) yield of S1 scintillation in DS-50 is 7.9 PE/keV \cite{g}. This project focuses on the scintillation S1 signal and its components. 

The S1 signal is generated as a result of two different Ar reactions. When an incoming particle interacts with an Ar atom in LAr, it causes the formation of both Ar excitons ($\text{Ar}^*$) and  Ar ions ($\text{Ar}^+$). The Ar excitons combine with the ground state Ar atom to form Ar excimers ($\text{Ar}_2^*$). The Ar ions combine with Ar atom to form charged dimers ($\text{Ar}_2^+$), capturing an electron through charge recombination to form highly excited Ar ($\text{Ar}^{**}$). These $\text{Ar}^{**}$ undergoes nonradiative de-excitation to $\text{Ar}^*$, and forms an Ar excimer ($\text{Ar}_2^*$) by combining with Ar atom. These Ar excimers formed via both Ar excitons and Ar ions in LAr are identified as singlet and triplet excimer states, which eventually dissociate to the ground state leading to the emission of scintillation light \cite{h}. 


Thus, the S1 scintillation exhibits a double-exponential decay form, with the main two components as singlet (fast emission) and triplet (slow emission). Ar dimers in the singlet and triplet states decay to the ground state within characteristic lifetimes; $\tau_s$ in the range of 2 ns -- 6 ns and $\tau_t$ in the range of 1100 ns -- 1600 ns \cite{c}. 

The probability of the Ar atoms to get excited to either the singlet or the triplet states is determined by the ionization density in the particle tracks. If the ionization density is high, then the relative intensity of the triplet component decreases \cite{i}.

Impurities, such as $\text{N}_2$, can interact with the Ar dimers in the excited states and de-excite them without the emission of light. Since the decay from the triplet state is significantly longer than that from the singlet state, the Ar excimers in the triplet state are more affected by the contaminants \cite{c}. 
\begin{equation}
Ar_2^* + N_2 \longrightarrow 2 Ar + N_2 \,
\qquad
\end{equation}

The UAr drawn from the cryostat may contain chemical impurities at levels ranging from ppm to ppb levels. For maintenance purposes, the hot getter was bypassed in the circulation loop for 5 days. This might bring some contaminants in LAr, that could have an impact on the apparent lifetime of the S1 triplet component. This brings forth the interest in comparing the lifetimes of the triplet component during the getter-on and off times in DS-50. Although it is highly unlikely for nitrogen or any other impurities to be present inside the TPC after purification, it might be that gas is being trapped in valves and other parts of the lines in the bypass.

\section{Data selection of the S1 triplet component event-wise analysis}
The primary objective of this contribution is to examine if there is any degradation in the triplet lifetime during the getter-off time period. Considering the potential of $\text{N}_2$ contamination at the level of ppm to cause a significant reduction in the lifetime of the slow component based on literature findings, observing a decrease in triplet lifetime during the getter-off duration would suggest the possible presence of $\text{N}_2$ in the LAr in DS-50. 

In the event-wise analysis, waveforms of individual S1 events in DS-50, recorded using the PMTs, are analyzed for the whole three year time period. A similar analysis has already been done to check the triplet lifetime over three years, by fitting the average waveform \cite{j}. The advantage of fitting event by event is that the average waveform might include abnormal events such as pileup events. In addition, fitting each waveform separately accounts for individual variations, unlike the average waveform, thus providing a more in-depth understanding of the data. 

Since information on the events is available for the pre-processed data, suitable event selections are applied to obtain "good" events. The following criteria are applied in this analysis for the event selection:
\begin{enumerate}
    \item $\Delta t > 20\times10^{-3}$ s : The time between 2 consecutive S1 events.
    \item $0.1 < \text{s}1_{f_{90}} < 0.5$ : The factor $f_{90}$ is the ratio between the first 90 $ns$ of the S1 pulse and the total S1 pulse (integral of S1 pulse over 7 $\mu s$). Only ER events are required for this study. The $\text{s}1_{f_{90}}$ factor for ER ($\beta$-induced) events is around 0.3.
    \item $\text{s}2 > 0$ : Events with both S1 and S2.
    \item $t_{drift} > 20$ $\mu s$: Tdrift is the time between S1 and S2 signals. To avoid overlapping between the two signals, this cut is applied.
    \item 1000 PE $< \text{s}1 < 1600$ PE : Lower limit to select only those events with higher statistics and upper limit is to avoid saturation of the signal.
    \item $\text{s}1_{fwhm} < 0.1$ $\mu s$ : To avoid events that are caused by pile up in S1 and S2.
    \item \(\lvert \text{s1}_{\text{tba}} \rvert < 0.9\) : The top-bottom asymmetry (TBA) parameter measures the difference between the PEs detected at the top and bottom of the PMTs relative to their total. This factor should be less than 1 to obtain better events from the center of the fiducial volume.
\end{enumerate}

\section{The waveform fit functions and the method of event-wise analysis}

The waveforms obtained after applying the cuts are grouped at 1000 events per run from the whole dataset (except getter-off days, for which all events in each run are considered). These waveforms are fitted using a full response function that consists of three different contributions \cite{k}. They are as follows; \\
The Ar scintillation time profile, which is described by;
\begin{equation} \label{eq:myequation1}
\begin{aligned}
F(t,\tau_s,\tau_t,p_s) \, &= \, \frac{p_s}{\tau_s} e^{-\frac{t}{\tau_s}} + \frac{1-p_s}{\tau_t}e^{-\frac{t}{\tau_t}}
\end{aligned}
\end{equation}
where $p_s$ and $p_t$ are the probabilities to populate the singlet and triplet states, respectively, and $\tau_s$ and $\tau_t$ are the decay times of the singlet and triplet states, respectively.\\
The TPB (TetraPhenyl Butadiene) re-emission contribution, given by
\begin{equation} \label{eq:myequation2}
\begin{aligned}
H(t,\tau_{TPB},p_{TPB}) \, &= \, (1 - p_{TPB}) + \frac{p_{TPB}}{\tau_{TPB}}e^{-\frac{t}{\tau_{TPB}}}
\end{aligned}
\end{equation}
Here, $p_0 = 1-p_{TPB}$ is the amplitude of prompt TPB re-emission component.\\
TPB is a wavelength shifter used in DS-50. It has one prompt ($p_0$) and more than one delayed re-emission components. In this analysis, only one delayed component ($p_{TPB}$) is considered for TPB.\\
The detector response, characterized by a Gaussian distribution with its peak at $\text{t}_0$, which is the reference time, and the resolution $\sigma$, is given by,
\begin{equation} \label{eq:myequation3}
\begin{aligned}
G(t',\sigma) \, &= \, \frac{e^{-\frac{t'^2}{2\sigma^2}}}{\sqrt{2\pi\sigma^2}}
\end{aligned}
\end{equation} 
where $t' = t-t_0$.\\
The convolution of these three functions is associated with the pulse height ($\textit{A}$) to obtain the full response function,
\begin{equation} \label{eq:mytequation4}
\begin{aligned}
P(t,\theta,t_0,A) \, &= \, A \times R(t',\theta)
\end{aligned}
\end{equation}
where $\theta$ is the set of parameters including $p_s, p_t, \tau_s \text{ and } \tau_t$,
\begin{equation} \label{eq:myequation5}
\begin{aligned}
R(t,\theta) \, &= \, F(t,\tau_s,\tau_t,p_s) \, \circledast \, H(t,\tau_{TPB},p_{TPB}) \, \circledast \, G(t,\sigma)
\end{aligned}
\end{equation}
The triplet (slow) component of the waveform is fitted also using an exponential function, given by
\begin{equation} \label{eq:myequation6}
\begin{aligned}
Y(t, \tau_t, A) &= \frac{A}{\tau_t} e^{\left(-\frac{t}{\tau_t}\right)}
\end{aligned}
\end{equation} 

\begin{figure}[htbp]
\centering
\includegraphics[width=.6\textwidth]{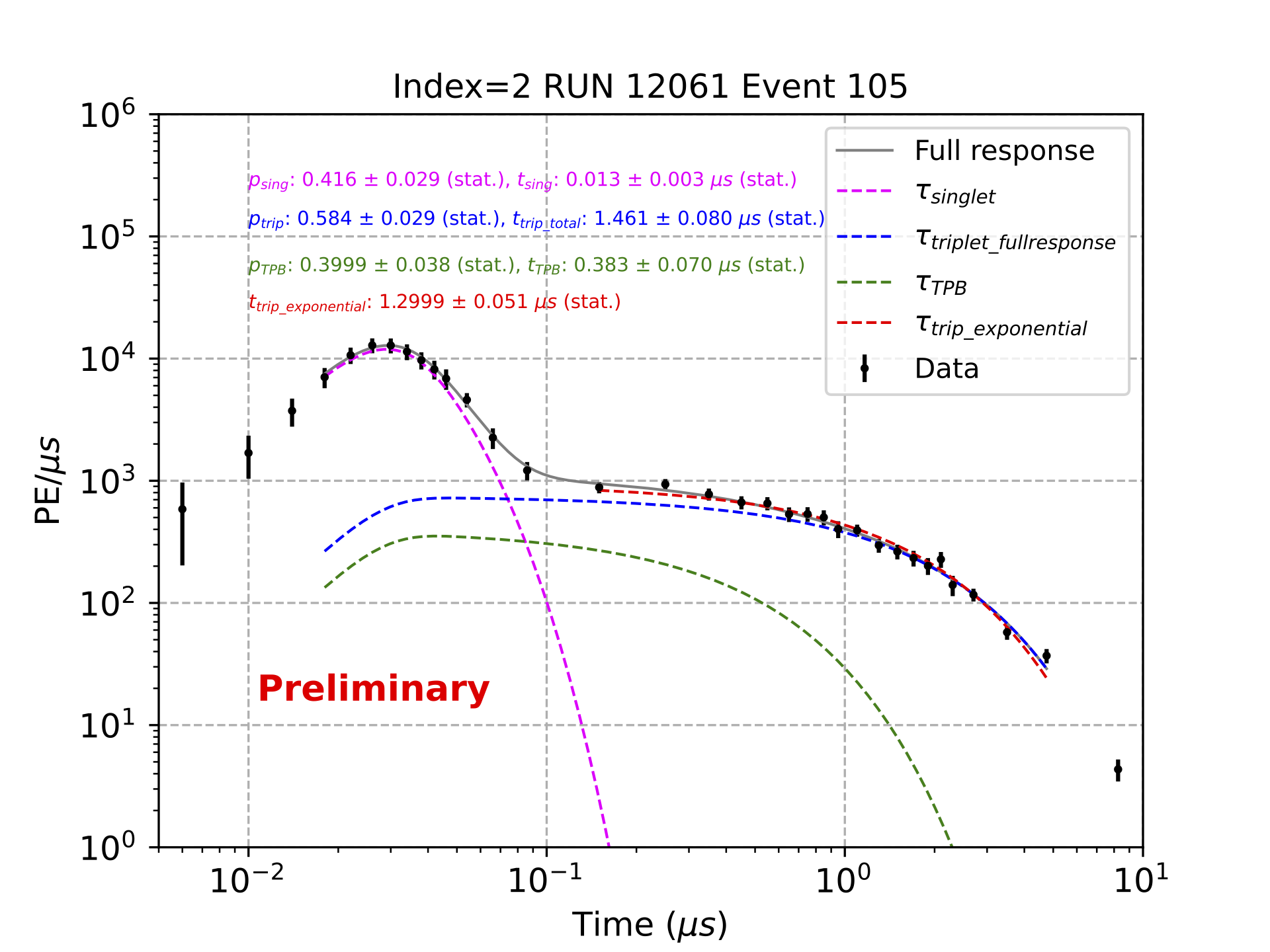}
\qquad
\caption{An example of an individual waveform, after selection cuts, with more than  10 PE/bin, fitted for the full response fn. (grey), and components (dotted lines); singlet (magenta), triplet (blue) scintillation, and TPB (green) re-emission components. An exponential fn. (red) is also shown. \label{fig:i}}
\end{figure}

The same data and selection cuts are used in the average waveform analysis \cite{j}, except for the S1 range in [100, 20000] PE. In the average waveform analysis, these selected waveforms are grouped into sets of 50000 PE, and the average and error on the mean are calculated for each group. Using the full response function, a $\chi^2$ fitting is performed on each average waveform over the range [0.01, 10] $\mu\text{s}$, and associated an average time of each run to each fitted waveform. The average waveform analysis reported a weighted mean triplet lifetime of $1.375 \pm 0.001 \, \mu s$ considering the errors, over the three years, and no reduction of lifetime during the getter-off period.

\begin{figure}[htbp]
\centering
\includegraphics[width=.6\textwidth]{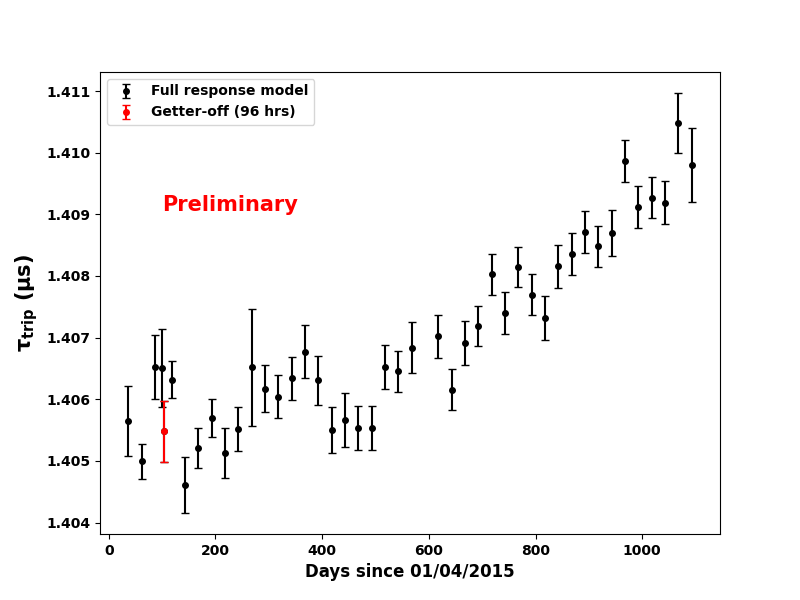}
\qquad
\caption{The triplet lifetime, from fitting individual waveforms with full response fn., over three years. Triplet lifetime value during the getter-off period is shown in red.\label{fig:j}}
\end{figure}

In the event-wise analysis, a $\chi^2$ fitting is performed on each individual waveform using the full response fit function given in eq. \eqref{eq:mytequation4} and an exponential function given in eq. \eqref{eq:myequation6}. A typical waveform fitted with the full response function, along with the additional exponential function (red curve) fitted on the triplet component of the waveform, is shown in figure~\ref{fig:i}. The full response function is fitted to the multi-components over the range [0.018, 4.75] $\mu\text{s}$ and the exponential function is fitted over the range [0.15, 4.75] $\mu\text{s}$. In order to enhance statistics, the time bins were re-sized enough to accommodate at least 10 PE per bin. The error on each waveform comes from statistical uncertainties. All the events are then associated with their corresponding days of recording, and then grouped into bins of days. The weighted average and the error on the mean is computed for each bin of days. Figure~\ref{fig:j} shows the weighted average of each day-bin plotted against the midpoint of the day-bins for three years since the start of data taking (01/04/2015). The mean triplet lifetime value for the getter-off bin (only 4 day-bins considered, to obtain pure getter-off data) is shown in red.

The weighted mean of the triplet lifetime over three years in DS-50 using the event-wise analysis is observed to be $1.4070 \pm 0.00006 \, \mu s$. A reduction in triplet lifetime of around $0.2 \, \mu s$ was expected during the getter-off period for around $1$ ppm of $\text{N}_2$ contaminant \cite{c}. However, the triplet lifetime during the getter-off period did not show any degradation, which may imply the absence of $\text{N}_2$ contaminant greater than $1$ ppm in the LAr. In addition to this, an overall increase in the triplet lifetime in the order of $0.4\%$ is observed after the getter was re-installed.

\section{Conclusions}

The purity of Ar used in LAr TPC is of paramount importance, hence a dedicated event-by-event analysis is put forth to assess the purity of UAr used in the DS-50 experiment. The lifetime of the triplet component in DS-50 over three years is obtained as $1.4070 \pm 0.00006 \, \mu s$ using event-wise analysis. The inconsistency between the mean triplet lifetime values between event-wise and average waveform analysis could be from the uncertainties in the fit models. The uncertainty quoted here is statistical only. The systematic uncertainties are currently under evaluation. As shown in figure~\ref{fig:j}, the triplet lifetime during the getter-off period is consistent within uncertainties with the lifetime before and after this period. This may imply that if the LAr is $\text{N}_2$-contaminated, then the concentration of $\text{N}_2$ in DS-50 is less than 1 ppm. This also implies that either $\text{N}_2$ may not be the cause of SEs, or SEs may be caused by less than 1 ppm concentration of $\text{N}_2$ contaminant.


\acknowledgments

The author is supported by the National Science Centre, Poland
(2021/42/E/ST2/00331). The speaker acknowledges support from the EU's Horizon 2020 research and innovation programme under grant agreement No 952480 (DarkWave project) and the International Research Agenda Programme AstroCeNT (Grant No. MAB/2018/7) funded by the Foundation for Polish Science from the European Regional Development Fund.





\end{document}